\documentclass[aps,preprint,noshowpacs,nofootinbib,superscriptaddress,groupedaddress, floatfix,superscriptaddress, preprintnumbers]{revtex4-1}  
 
\usepackage{graphicx} 
\usepackage{dcolumn}   
\usepackage{bm}        
\usepackage{amssymb}   
\usepackage[colorlinks=true]{hyperref}

\usepackage{multirow}
\usepackage{feynmf}
\usepackage{amsmath,amssymb}

\usepackage{graphicx,bm}

\usepackage{slashed}

\usepackage{natbib}

\usepackage{epstopdf}

\usepackage{ulem} 

\usepackage[usenames]{color}

\usepackage{float}

\renewcommand\sout{\bgroup \color{red} \ULdepth=-.5ex \ULset}

\newcommand{\be}{\begin{equation}}
\newcommand{\ee}{\end{equation}}
\newcommand{\ba}{\begin{eqnarray}}
\newcommand{\ea}{\end{eqnarray}}

\usepackage{soul} 

\newcommand{\bea}{\begin{eqnarray}}
\newcommand{\eea}{\end{eqnarray}}
\def\s{\sigma}
\def\th{\theta}
\def\m{\mu}\def\n{\nu}\def\w{\omega}
\def\bfk {{\bf k}}\def\bfq {{\bf q}}\def\d{\delta}
\newcommand{\eq}[1]{Eq.~(\ref{#1})}
\begin{document}
\begin{flushright}
\end{flushright}


\title{Might Normal Nuclear Matter be Quarkyonic? }

\author{Volker Koch}
\affiliation{Nuclear Science Division, Lawrence Berkeley National Laboratory, 1 Cyclotron Road, Berkeley, CA 94720, USA}
\email{vkoch@lbl.gov}

\author{Larry McLerran }
\affiliation{Institute for Nuclear Theory, University of Washington, Box 351550, Seattle, WA 98195, USA}
\email{mclerran@me.com}

\author{Gerald A. Miller}
\affiliation{Department of Physics, University of Washington, Seattle, WA 98195, USA}
\email{miller@uw.edu}

\author{Volodymyr Vovchenko}
\affiliation{Department of Physics, University of Houston, 3507 Cullen Blvd, Houston, TX 77204, USA}
\email{vvovchenko@uh.edu}

\date{\today}

\begin{abstract}
The possibility that nuclear matter might be Quarkyonic is considered.
Quarkyonic matter is high baryon density matter that is confined but can be approximately thought of as a filled Fermi sea of quarks surrounded by a shell of nucleons.  
Here, nuclear matter is described by the IdylliQ sigma model for Quarkyonic matter, generalizing the non-interacting IdylliQ model~[Y. Fujimoto et al., \href{https://doi.org/10.1103/PhysRevLett.132.112701}{{Phys. Rev. Lett. {\bf 132}, 112701 (2024)}}] to include interactions with a sigma meson and a pion.  
When such interactions are included, we find that isospin-symmetric nuclear matter binds, with acceptable values of the compressibility and other parameters for nuclear matter at saturation.
The energy per nucleon and sound velocity of such matter is computed, and the isospin dependence is determined. 
Nuclear matter is formed at a density close to but slightly above the density at which Quarkyonic matter forms.  
Quarkyonic matter predicts a strong depletion of nucleons in normal nuclear matter at low momentum.  
Such a depletion  for nucleon momenta  $k \lesssim 120$~MeV is shown to be consistent with electron scattering data.
\end{abstract}


\maketitle

\section{Introduction}

In Ref.\cite{McLerran:2007qj}, the authors argued that in the limit of a large number of colors, $N_c$, matter at high baryon density consists of the Fermi sphere of quarks surrounded by a small momentum shell of baryons. 
Such matter is called Quarkyonic since both nucleon and quark degrees of freedom are important. 
It has been argued~\cite{McLerran:2018hbz}, that Quarkyonic matter naturally explains the maximum of the density dependence of the speed of sound extracted from neutron star phenomenology \cite{Fujimoto:2019hxv,Tan:2020ics}. 
The existence of Quarkyonic matter is also supported by effective lattice theory for QCD with heavy quarks~\cite{Philipsen:2019qqm}.
Many dynamical models for Quarkyonic matter have been proposed \cite{Duarte:2020xsp,Jeong:2019lhv,Kojo:2021ugu,Sen:2020peq,Kovensky:2020xif,Cao:2020byn,Poberezhnyuk:2023rct}, mostly involving a quark-baryon mixture and repulsive excluded volume interactions for the baryons, which, however, might not result in Quarkyonic matter to be the state of lowest energy \cite{Koch:2022act}. 
This issue has been recently addressed in a novel construction for Quarkyonic matter -- the so-called IdylliQ model of Quarkyonic matter~\cite{Fujimoto:2023mzy}.

The IdylliQ model is a theory of matter made from free nucleons which are only subject to the Pauli exclusion principle of the nucleons {\it and} the quarks inside the nucleon. The nucleons are composed of quarks with a momentum space probability distribution, $\varphi(p)$, with the normalization
$\int ~\frac{d^3p}{(2\pi)^3} \varphi(p)=1$.
The momentum space distribution of the quarks, $f_Q(p)$ is then related to that of the nucleons, $f_N(q)$ via
\begin{equation}
   f_Q(p) = \int~ \frac{d^3k}{(2\pi)^3} \varphi(|p-k/N_c|) f_N(k) 
   \label{eq:rhoQ}
\end{equation}
where $N_c$ is the number of colors. Fermi statistics are imposed for both quarks and nucleons, $0 \le f_Q(p), f_N(k) \le 1$.  The IdylliQ model is a dual description of baryonic matter. Contrary to earlier implementations of Quarkyonic matter, where baryons  were assumed to co-exist on top of a filled Fermi-sphere of quarks, in the IdylliQ model, the matter may either be viewed in terms of baryon degrees of freedom or as quarks 

This theory has two phases.  
The low-density phase is an ideal gas of nucleons and is best viewed in terms of baryons.
The high-density phase is Quarkyonic. There the low momentum states of the quarks are fully occupied, $f_Q(p)=1$, up to a certain momentum, $p<p_{\rm bu}$ above which the phase space density decreases exponentially. In the dual picture of baryons, this implies that the nucleon density is composed of a bulk contribution of nucleons where at momentum below $k<N_c p_{\rm bu}$ the phase space density is $f_N = 1/ N_c^3$ and fully occupied, $f_N =1$,  in a thin shell up to the Fermi-surface of the baryons. 
The factor of $1/N_c^3$ for the nucleon occupation number at low momentum can be understood as follows: With $N_c$ quarks inside a nucleon, the typical momentum of a nucleon is $N_c$ times that of its quarks.  This means that the phase space for nucleons is $k^3/p^3 \sim N_c^3$ and by \eqref{eq:rhoQ},. Thus the occupation of baryons dual to a quark occupation of $f_Q=1$ is $f_N \sim 1/N_c^3$

The transition density between ordinary nucleonic matter and Quarkyonic matter is determined by the onset of the full occupation of quark states. 
This begins at quark momentum $k = 0$. 
Using Eq.\eqref{eq:rhoQ} the condition corresponds to
\begin{equation}
 1 = f_Q(0) =  \int_{k < k_F^{\rm crit}}~ {{d^3k} \over {(2\pi)^3} }~\varphi(k/N_c) f_N(k)
 \label{eq:onset_condition}
\end{equation}
which determines the critical Fermi momentum $k_F^{\rm crit}$ for quark saturation.

Before turning to the specific choice for the probability distribution, $\phi$, used in the IdylliQ model computation,  it is useful to understand some of the typical scales for nucleons in nuclear matter.  To this effect, we can consider a Gaussian distribution which would arise in a harmonic oscillator model for the quark wavefunction in a nucleon.  Such a distribution will underestimate the long distance tail of the wavefunction, but it will give us some semi-quantitative insight.  A generic feature of Quarkyonic matter is the formation of a filled Fermi sea of quarks at low momentum.  We will ask whether or not it is reasonable to expect that such a filled Fermi sea forms near nuclear matter density?  We parameterize the probability to find a quark inside of a hadron by Gaussian density
\begin{equation}
 \varphi_{\rm gauss}(p) = 8\pi^{3/2} R^3e ^{-p^2R^2}
\end{equation}
and then determine nucleon density for which quark density saturates through Eq.~\eqref{eq:onset_condition}. 
Note that the parameter $R$ defines the root-mean-square nucleon radius $r_{\rm RMS}$ as $R = \sqrt{2 \over 3} r_{\rm RMS}$.  
We find the critical Fermi momentum for quark saturation $k_F^{\rm crit} \equiv N_c p_F^{\rm crit}$ is comparable to the nuclear Fermi gas momentum
$k_F^{\rm NM} \simeq 265$~MeV for a reasonable choice of $r_{\rm RMS}$. Assuming a normal Fermi distribution, one finds that the right-hand-side of \eq{eq:onset_condition} is  given approximately by $0.41 ({k_F^{\rm NM}} r_{\rm RMS})^3$, so that the critical Fermi momentum, $k_F^{\rm crit} $, for $f_Q(0)$ to be unity  is found to be
 $k_F^{\rm crit} \sim 1.01 ~k_F^{\rm NM}$ for $r_{\rm RMS} \sim 1$~fm, and $k_F^{\rm crit} = 1.3~k_F^{\rm NM}$ for $r_{\rm RMS} \sim 0.8$~fm.
These numbers show that normal nuclear matter is very close to the critical density at which Quarkyonic matter might form in a nucleus, and whether this ultimately takes place lies in the details of the specific implementation of the IdylliQ concept.    
Perhaps it is not an accident that the density of nuclear matter is close to that of quark saturation and that Quarkyonic matter plays a role in nuclei?

The simplest form of the IdylliQ model uses a specific choice of probability distribution,
\begin{equation}
  \varphi(p) = {{2\pi^2} \over \Lambda^2} {{e^{-\mid p \mid /\Lambda}} \over {\mid p \mid}} 
\label{eq:quark_dist}
\end{equation}
for which the theory is analytically solvable.  Here, $\Lambda$ is the typical momentum scale of a nucleon inside a hadron and should be of the order of the QCD scale, $\Lambda_{\rm QCD}$.

 The singular factor of $1/p$ allows for an analytically solvable model, as we will see below.  However, its origin might be more general.  If one includes a pion cloud around nucleons, this pion cloud will generate a logarithmically divergent charge radius in the limit of zero pion mass~\cite{Beg:1972vyx}.  A factor of $1/p$ in the distribution of quarks reproduces this divergence.  If this divergence appears in the computation of a physical process, we should remember to cut it off at the pion mass scale.  A proper treatment of the pion cloud awaits further analysis.  In our computations of nuclear matter below, no infrared divergence arises, and the theory without a pion cloud will be sufficient for our crude semi-quantitative analysis, and provides us with analytic insights which will be useful in a more developed theory.

In the integral for the onset condition, Eq. \ref{eq:onset_condition}, it is a good approximation at large $N_c$ to take $\varphi(k/N_c) \simeq \frac{2 \pi^2}{\Lambda^2} \frac{N_c}{k}$, so that Eq. \ref{eq:onset_condition} reads
\begin{equation}
   1 = \frac{N_c}{\Lambda^2} \int~kdk~ f_N(k),
\end{equation}
and the critical Fermi momenta for nucleons  for the onset of Quarkyonic matter is
\begin{equation}
   k_F^{\rm crit} = \sqrt{2 \over N_c} ~\Lambda
\end{equation}
The $N_c$ dependence of this equation requires that the density of the transition to Quarkyonic matter is
\begin{equation}
 n_{\rm onset} \sim {\Lambda^3 \over N_c^{3/2}}
\end{equation}

The $N_c$ dependence of the onset density is quite remarkable.  Its value is parametrically lower than the natural QCD scale, $\Lambda^3_{\rm QCD} \sim \Lambda^3$.  At such a parametrically lower density, interactions might be expected to be weaker than their natural QCD scale, which in the naive large $N_c$ limit would give binding energies of order the nucleon mass, $m_N \sim N_c \Lambda$.  Since the binding energy of nuclear matter is much less than the nucleon mass scale, obtaining in a simple and natural way such weak binding has been a persistent problem of the large $N_c$ limit applied to nuclear matter.    Such a parametrically low density for the formation of Quarkyonic matter suggests that such matter might be important for nuclear matter, or that nuclear matter itself might be Quarkyonic.
We note that such a possibility has also been suggested in Ref.~\cite{Philipsen:2019qqm} based on QCD with heavy quarks.

To determine whether it might be possible that the onset of Quarkyonic matter is related to that of nuclear matter, we need a theory that includes the interactions between nucleons.  If the matter is at low density, as suggested above, then an approximation to the theory is the sigma model where interactions are included by pions, sigma mesons, and nucleons.  We will take this theory to be in motivated by the linear sigma model in that we use the same coupling for the pions and the sigma mesons.

First, we recall that in the Walecka mean field model \cite{Serot:1984ey}, nuclear binding in isospin symmetric matter results from the trade-off between a scalar interaction which is attractive, and a repulsive vector interaction. 
The scalar interaction energy is proportional to the square of the scalar density, $n_{s}$,  with
\begin{equation}
   n_{s} =  4 \int~ {{d^3k} \over {(2\pi)^3} }  {m \over E} \Theta(k_F-k).
\end{equation}
The vector density, $n_v$, on the other hand, is given as 
\begin{equation}
n_{v} =  4 \int~ {{d^3k} \over {(2\pi)^3} }  \Theta(k_F-k)
\end{equation}
We see that the scalar density weakens relative to the vector density as the Fermi momentum increases, and this can generate a binding potential.

In contrast to the Walecka model, there is no vector interaction in the IdylliQ sigma model.  
Instead, the repulsion corresponding to the vector interactions in the Walecka model arises from the Fermi exclusion principle in the quark sector and the associated energy cost to fill higher momentum states in order to have the same baryon density.  
This {\it Fermi repulsion} must be what generates the repulsive core responsible for binding.  
The attraction is generated by the scalar meson interaction term.  
Notice that a scalar density goes like $n_{s} \sim k_F^2$ at large $k_F$  so that a scalar interaction $n_{s}^2/M_{s}^2 \sim k_F^4$ at large $k_F$.  
This is the behavior of a scale-invariant system.  
Such an interaction term will not necessarily overpower the kinetic energy of the nucleons at high momentum, which is also scale invariant.  
On the other hand, a vector interaction would go like $k_F^6$ at high momentum and thus violate the approximate scale invariance of QCD at high density.

The essential feature of Quarkyonic matter that needs experimental testing is the formation of a filled Fermi sea of quarks. 
If the momentum characteristic of the filled quark Fermi sea is $p_Q$, then the typical momentum for nucleons
is $k_N \sim N_c p_Q$.  
Recalling that the  quark baryon number is $1/N_c$, as well as the color degeneracy of $N_c$, one obtains that the total baryon number density in the filled Fermi sea of quarks is 
$n_B \sim {1 \over N_c} N_c k_Q^3 \sim p_Q^3 \sim k_N^3/N_c^3$, i.e. $1/N_c^3 \sim 3-4$\% of the  expected value.  
Therefore, there is, for all practical purposes, a ``hole" at the bottom of the baryon Fermi Sea.

Surely if there is a hole at the bottom of the nucleon Fermi sea, one might have expected to see its effect in electron scattering.  Such scattering provided some of the best measurements of Fermi momenta and properties of nucleon momentum distributions in nuclei~\cite{Sick:1970ma},\cite{Day:1987az},\cite{Day:1989nw}, through a comparison with the corresponding theoretical computations~\cite{Moniz:1969sr,Moniz:1971mt}.
In Ref.\cite{Day:1989nw}, in particular, electron scattering was analyzed in an energy range that is sensitive to the distributions of nucleons inside of nuclei with electron energy of $500$~MeV and $60^\circ$ scattering angle.  
The data were Coulomb corrected and extrapolated to infinite nuclear size so as to represent infinite nuclear matter.  
Later in this paper, we compare the results of our computations within Quarkyonic matter and argue that the data are suggestive of a strong depletion of the Fermi sea of nucleon at momenta $k \lesssim 120$~MeV.
The computation we present for nuclear matter suggests a larger value of the characteristic momentum of the hole $k \sim 120-180$~MeV, which is currently still subject to large model uncertainties.

Of course,  the IdylliQ model is not yet sufficiently well developed to expect a precise agreement between the parameters of the IdylliQ model and that of the hole in the Fermi sea that is perhaps seen in the electron scattering data.  
However, we will see that the Quarkyonic model predicts the correct semi-quantitative features of nuclear matter, and the hole in the Fermi distribution that describes the electron scattering data. 
There is much that can be done to refine the theory, but it is remarkable that the region where Quarkyonic matter first appears might be close to and slightly below the density of nuclear matter.  
Nevertheless, the hypothesis that Quarkyonic matter may play a role in conventional nuclei is a radical one, and may very well be possible to rule out by further theoretical computation and comparison with experimental results, and we view the results we present as very speculative and tentative

The outline of this paper is as follows:  In Sec.~\ref{sec:idylliqsigma}, we introduce IdylliQ Sigma model for nuclear matter.
We review the computations of the mean field and exchange energy for sigma mesons and pions, and the effective nucleon mass.  
In Sec.~\ref{sec:eos} we present computations of the energy per baryon and argue that nuclear matter exists at a density slightly above that of the transition to Quarkyonic matter. 
We compute the isospin dependence of the energy per baryon and the sound velocity.  
In Sec.~\ref{sec:elscat} we compute electron scattering from a Quarkyonic momentum distribution of nucleons, and show that the resulting distribution has properties that describe the observed scattering if we have a depletion in the nucleon momentum distribution below $k \lesssim 120$ MeV. 
In the final section~\ref{sec:summary}, we discuss the many limitations and possible paths for improvement of our work.

\section{IdylliQ Sigma model}
\label{sec:idylliqsigma}

We begin computations by first using the IdylliQ model to compute the dependence of the differential nucleon number density as a function of momentum for a fixed total nucleon density. 
The computation method is as in Ref. \cite{Fujimoto:2023mzy}.  
We will consider various possibilities for the isospin to baryon number density ratio but will be most interested in the iso-singlet matter.  
This computation will determine the momentum space shell structure of the nucleon number distribution. When this is done, we will compute the effect of sigma meson and pion exchange.  
The sigma mesons will contribute a mean-field term, and there will additionally be exchange energy terms for both the sigma meson and pion.  
We will present computations for a fixed nucleon or constituent quark mass, not accounting for modifications associated with the sigma meson condensate.  
We also consider a density-dependent mass but it does not change the picture appreciably except for the case of a larger mass sigma field, where the compressibility is increased, and the lower momentum of the shell is decreased.  
This demonstrates that the value of the lower momentum of the shell will be sensitive to the stiffness of the equation of state.  
It turns out that the equation of state for neutron matter with a fixed nucleon mass in our computations is softer than what is favored for neutron stars, with the sound velocity squared never exceeding 1/3 in the range of densities appropriate for neutron stars.  
The case with a density-dependent mass gives somewhat better agreement with observations with the sound velocity square slightly exceeding 1/3 at the highest densities. Probably the equation of state needs to be a little stiffer, and no doubt our approximation of only including the lowest-order contribution of the sigma meson and pions is not quite correct.    
This might explain the somewhat small value of the lower momentum of the nucleon shell found by comparison with the electron scattering data.

\subsection{Shell structure}

In the following we discuss the properties of the IdylliQ model in terms of its  baryon picture, as this is more suitable to introduce interactions and also analyse predictions for electron scattering. The baryon density is expressed as 
\begin{equation}
n_B = \frac{\gamma_s}{2\pi^2} \int dk \, k^2 \, [f_p(k) + f_n(k)],
\end{equation}
where $\gamma_s = 2$ is the spin degeneracy factor, and $f_{p,n}(k)$ are the occupation numbers for protons and neutrons, respectively.

We implement the Pauli exclusion principle among quarks following the exactly solvable IdylliQ Quarkyonic matter model of~\cite{Fujimoto:2023mzy}.
The shell structure evolves with baryon density, which we take to be unaffected by the interactions we consider here.
We reiterate below the key features of the resulting shell structure, both for symmetric and pure neutron matter.

\subsubsection{Symmetric nuclear matter}

The solution found in Ref. \cite{Fujimoto:2023mzy} has an 
occupation number densities for protons and neutrons (see Eq. (12) of \cite{Fujimoto:2023mzy})
\begin{equation}
f_p(k) = f_n(k) = \frac{1}{N_c^3} \Theta(k_{\rm bu} - k) + \Theta(k_{\rm sh} - k) \Theta(k - k_{\rm bu})~.
\label{eq:piecewise}
\end{equation}
This shell-like structure of the baryon density, in turn, can be used to compute the quark density.  To see how this solution arises, we first use Eq.~\eqref{eq:rhoQ} and Eq.~\eqref{eq:quark_dist}, to derive 
\begin{equation}
  \left\{ -\nabla_k^2 + {1 \over \Lambda^2} \right\} f_Q(p_Q) = {N_c^3 \over \Lambda^2} f_N(N_c p_Q)
  \label{eq:differential_fQ}
\end{equation}
to determine the quark density from the nucleon density.  
The solution to minimizing the energy density is found by having either $f_Q = 1$ or $f_N = 1$.  
This is plausible since if a low energy state is available to be occupied it will fill up before filling up a higher energy state.  
However, both the nucleon and quark distribution cannot simultaneously saturate at unity in overlapping regions of momentum ($k_Q = k_N/N_c$) without violating Eq.~\eqref{eq:differential_fQ}.
Indeed, if both are constant, this equation requires $f_N = {1 \over N_c^3} f_Q$ for $k_N = N_c p_Q$.  

Instead, one considers a piece-wise solution where these conditions are satisfied. 
At low densities, this is simply the ideal Fermi gas where $f_N = \theta(k_F-k)$, i.e. $k_{\rm bu} = 0$ and $k_{\rm sh} = k_F$.  
This solution no longer works for $k_F > k_{F}^{\rm crit}$ because one will find that at low momentum $f_Q \ge 1$.  
Instead, above $k_{F}^{\rm crit}$, we construct a solution such as given by Eq.~\eqref{eq:piecewise}.  
If we fix the upper momentum of the shell, $k_{\rm sh}$, then the lower momentum, $k_{\rm bu}$ is determined by solving Eq.~\eqref{eq:differential_fQ} and requiring that $f_Q(p) = 1$ for $p \le k_{\rm bu}/N_c$.
The solution Eq.~\eqref{eq:differential_fQ} for the quarks is a filled Fermi sea for momentum $p_Q < k_{\rm bu}/N_c$ with an exponentially falling profile above this momentum.  
See Ref. \cite{Fujimoto:2023mzy} for the details of this construction.
The value of the upper shell momentum $k_{\rm sh}$ is determined to reproduce a given baryon density.

\subsection{Pure neutron matter}

When we have neutrons $(udd)$ only, we have the saturation of $d$ quark sea, while $u$ quark levels are half-filled. We have
\begin{align}
f_p(k) & = 0, \\
f_n(k) & = \frac{3}{2} \, \frac{1}{N_c^3} \Theta(k_{\rm bu} - k) + \Theta(k_{\rm sh} - k) \Theta(k - k_{\rm bu})~.
\end{align}

For each baryon momentum state in neutron matter, we have 2 neutrons and thus four down quarks, whereas in nuclear matter we have two additional protons and thus six down quarks. Therefore, in order to saturate the down-quark Fermi distribution at $f_d(q = 0) = 1$ the additional factor of $3/2$ is needed.
As in symmetric matter, $k_{\rm bu}$ and $k_{\rm sh}$ are functions of $n_B$ only.

\subsection{Sigma model interactions}

In order to describe nuclear matter we need to include interactions. 
Specifically, we 
consider mean field, $\sigma$ and $\pi$ exchange energy contributions to the energy density. 
We explore various scenarios, most of which lead to the same qualitative behavior for the equation of state and speed of sound. 

Here, we will first omit the effect of density dependence of the nucleon mass and set it to vacuum value. 
We consider the effects of the density dependence below.

The energy density contains kinetic, mean-field, and exchange energy contributions:
\begin{equation}
\varepsilon(n_B) = \varepsilon_K + \varepsilon_{\rm MF} + \varepsilon_{\rm exch}^{\sigma} + \varepsilon_{\rm exch}^{\pi}.
\end{equation}
Here
\begin{align}
\label{eq:eK}
\varepsilon_K & =  \frac{\gamma_s}{2\pi^2} \int dk \, k^2 \, \sqrt{k^2 + m_N^2} \, [f_p(k) + f_n(k)], \\
\varepsilon_{\rm MF} & =  -\frac{g_s^2}{2 m_{\sigma}^2} n_{s}^2,
\end{align}
where $n_s$ is the scalar density
\begin{align}
\label{eq:rhosig}
n_s = \frac{\gamma_s}{2\pi^2} \int dk \, k^2 \, \frac{m_N}{\sqrt{k^2 + m_N^2}} \, [f_p(k) + f_n(k)].
\end{align}

The exchange energy terms read~\cite{Hu:2007na}
\begin{align}
\label{eq:exsig}
\varepsilon_{\rm exch}^{\sigma} & = \frac{\gamma_s g_s^2}{4 m_N^2} \int \frac{d^3q}{(2\pi)^3} \frac{m_N}{\omega_q} \int \frac{d^3p}{(2\pi)^3} \frac{m_N}{\omega_p} [f_p(q) f_p(p) + f_n(q) f_n(p)] \nonumber \\ 
& \quad \times \frac{\epsilon_q \epsilon_p - \mathbf{q} \cdot \mathbf{p} + m_N^2}{2(\epsilon_q \epsilon_p - \mathbf{q} \cdot \mathbf{p} - m_N^2) + m_\sigma^2}~ \nonumber \\
& = \frac{\gamma_s g_s^2}{32 \pi^4 m_N^2} \int d q \, q^2 \frac{m_N}{\omega_q} \int dp \, p^2 \frac{m_N}{\omega_p} [f_p(q) f_p(p) + f_n(q) f_n(p)] \nonumber \\ 
& \quad \times \int_{-1}^1 dx \frac{\epsilon_q \epsilon_p - qpx + m_N^2}{2(\epsilon_q \epsilon_p - qpx - m_N^2) + m_\sigma^2}
\end{align}
and
\begin{align}
\label{eq:expi}
\varepsilon_{\rm exch}^{\pi} & = \frac{\gamma_s g_s^2}{4 m_N^2} \int \frac{d^3q}{(2\pi)^3} \frac{m_N}{\omega_q} \int \frac{d^3p}{(2\pi)^3} \frac{m_N}{\omega_p} [f_p(q) f_p(p) + f_n(q) f_n(p) + 4 f_p(q) f_n(p)] \nonumber \\ 
& \quad \times \frac{\epsilon_q \epsilon_p - \mathbf{q} \cdot \mathbf{p} - m_N^2}{2(\epsilon_q \epsilon_p - \mathbf{q} \cdot \mathbf{p} - m_N^2) + m_\pi^2}~ \nonumber \\
& = \frac{\gamma_s g_s^2}{32 \pi^4 m_N^2} \int d q \, q^2 \frac{m_N}{\omega_q} \int dp \, p^2 \frac{m_N}{\omega_p} [f_p(q) f_p(p) + f_n(q) f_n(p) + 4 f_p(q) f_n(p)] \nonumber \\ 
& \quad \times \int_{-1}^1 dx \frac{\epsilon_q \epsilon_p - qpx - m_N^2}{2(\epsilon_q \epsilon_p - qpx - m_N^2) + m_\pi^2}
\end{align}
Here $\epsilon_{q} = \sqrt{m_N^2 + q^2}$.

\section{Equation of state of nuclear matter}
\label{sec:eos}

\subsection{Results with Fixed Nucleon Mass}

We consider calculations 
for two scenarios regarding the scalar meson mass, light~($m_\sigma = 0.5$~GeV) and heavy~($m_\sigma = 0.9$~GeV).
We also explore different scenarios where we only include the mean-field energy term~($\varepsilon_{\rm MF}$),  mean-field and scalar exchange terms~($\varepsilon_{\rm MF}$ and $\varepsilon_{\rm exch}^{\sigma}$), mean-field, scalar and pion exchange terms~($\varepsilon_{\rm MF}$, $\varepsilon_{\rm exch}^{\sigma}$, and $\varepsilon_{\rm exch}^{\pi}$).
The remaining model parameters, $g_s$ and $\Lambda$, are then fixed in each setup to reproduce the symmetric nuclear matter ground state, $E_{\rm bind} = \varepsilon /n_B - m_N = -16 \,\rm MeV$ at $n_B = n_0$, where $n_0=0.16\, \rm fm^{-3}$ is the ground state density of nuclear matter.

\begin{figure}[t]
    \includegraphics[width=.89\textwidth]{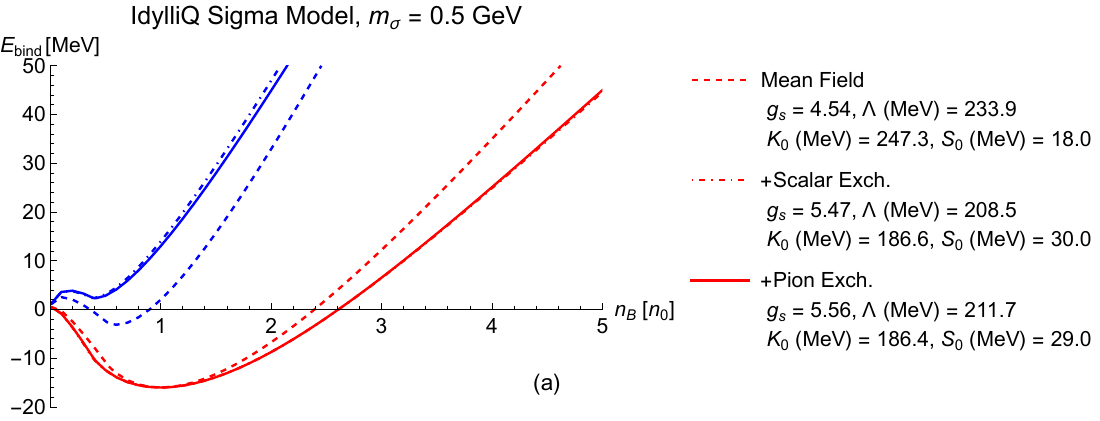}
    \vskip10pt
    \includegraphics[width=.89\textwidth]{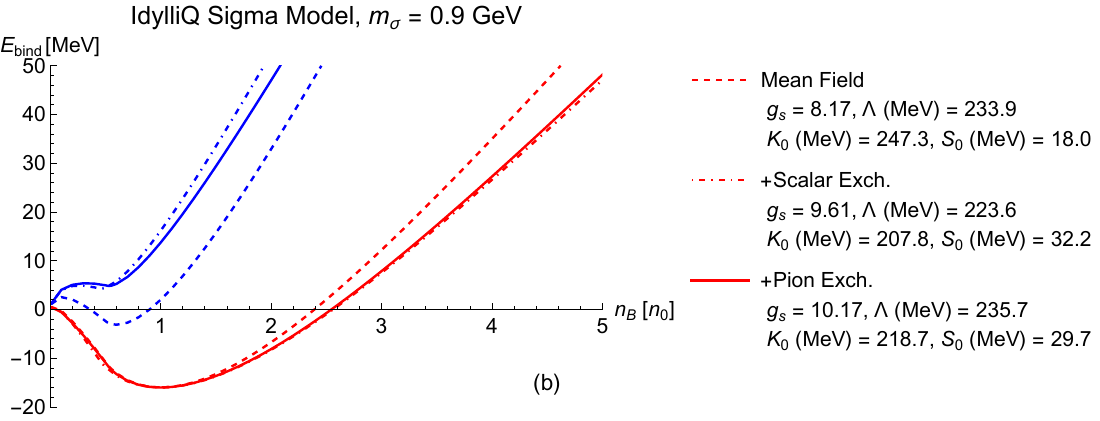}
    \caption{
    Baryon density dependence of the binding energy in the IdylliQ Sigma model in different configurations.
    The upper and left panels correspond to light~($m_\sigma = 500$~MeV/$c^2$) and heavy~($m_\sigma = 900$~MeV/$c^2$) scalar meson masses.
    Red and blue lines correspond to symmetric and pure neutron matter calculations, respectively.
    Different line styles correspond to different scenarios concerning the inclusion of scalar and/or pion exchange terms.
    }
    \label{fig:Ebind}
\end{figure}

\begin{figure}[t]
    \includegraphics[width=.49\textwidth]{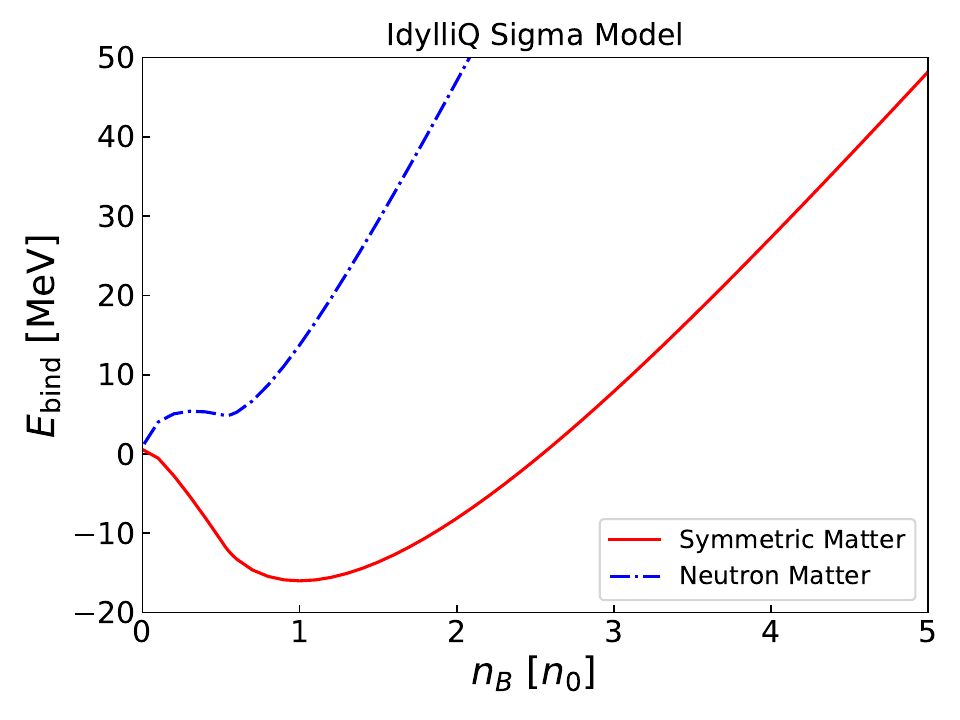}
    \includegraphics[width=.49\textwidth]{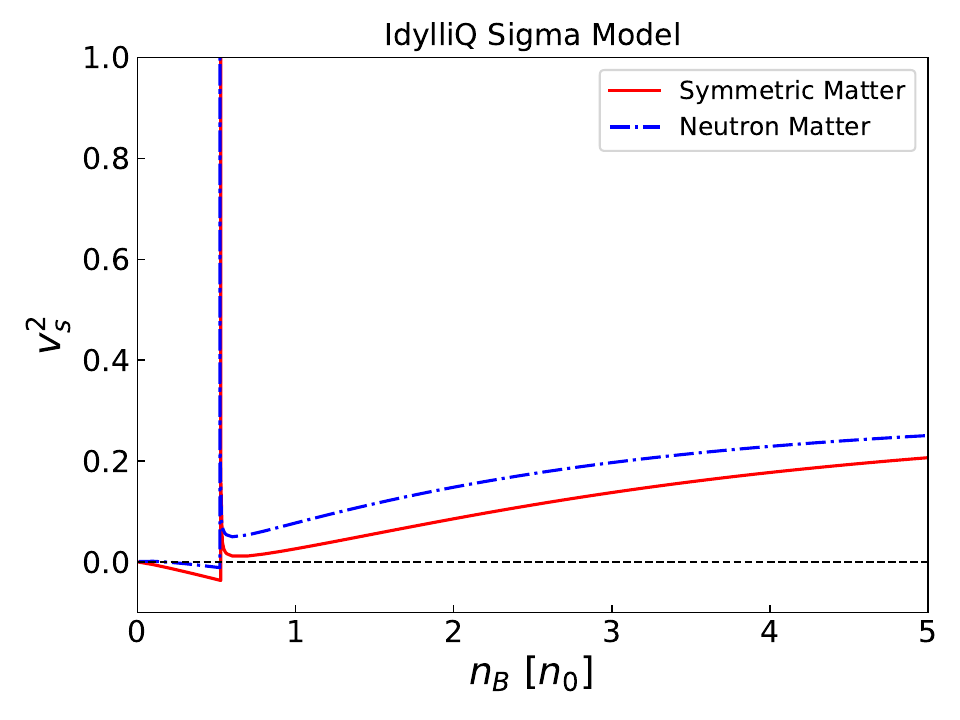}
    \caption{
    Binding energy~(left panel) and the squared sound velocity~(right panel) of symmetric nuclear~(solid red) and pure neutron~(dash-dotted blue) matter within the IdylliQ Sigma Model in our preferred scenario ($m_\sigma = 0.9$~GeV, scalar and pion exchange included, density-independent mass).
    }
    \label{fig:EoS}
\end{figure}

The results are depicted in Fig.~\ref{fig:Ebind} where we show the binding energy, $E_{\rm bind}$  as a function of the baryon density measured in units of the nuclear saturation density, $n_0$.  The results for symmetric nuclear matter are shown as red lines and that for neutron matter as blue lines. 
Calculations are performed on the mean-field level~(dotted lines), including scalar exchange term only~(dash-dotted lines), and scalar + pion exchange terms~(solid lines).
We find that the model predicts a reasonable value for the incompressibility, $K_0$, with $180 \, {\rm MeV} \lesssim K_0 \lesssim 270 \, \rm MeV$.
Also, the predicted values of the symmetry energy $S_0$ at $n_0$, are in a reasonable range ($29 \, {\rm MeV} \lesssim S_0 \lesssim 32 \, \rm MeV$) once iso-scalar and iso-vector exchange terms are included.
If the exchange terms are neglected~(scalar mean-field contribution only), the symmetry energy is too low~($S_0 \simeq 18$~MeV), and the model predicts bound neutron matter as seen by the dashed line in Fig.~\ref{fig:EoS}. 
Including the iso-scalar and iso-vector exchange terms is thus essential for a reasonable description of neutron matter in this framework.

\begin{figure}[t]
    \includegraphics[width=.49\textwidth]{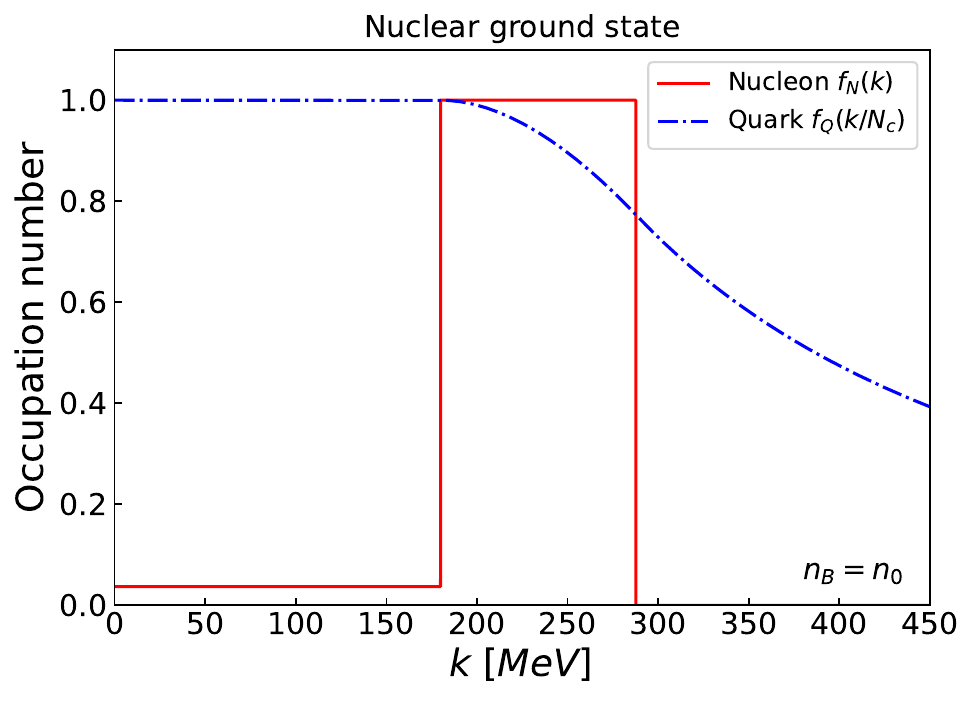}
    \caption{
    Occupation numbers of nucleons~(solid red line) and quarks~(dash-dotted blue line) as a function of the momentum $k$ in the nuclear ground state, $n_B = n_0$, in the same IdylliQ Sigma Model as in Fig.~\ref{fig:EoS}.
    Note that here, the quark momenta are scaled by the factor $N_c$. }  
    \label{fig:dist}
\end{figure}

The scenario with sigma and pion exchange and a sigma mass of $m_\sigma=0.9 \,\rm GeV$ and $g_s=10.1737$ is closest to the predictions of the sigma model where $g_s f_\pi \simeq m_N = 0.938$~GeV/$c^2$.
This is our preferred scenario, for which we have $K_0=219\,\rm MeV$ and $S_0= 29 \,\rm MeV$. 
In Fig.~\ref{fig:EoS}, we show the corresponding results for the speed of sound,
\begin{equation}
    v_s^2 = \frac{n_B}{\mu_B} \frac{d \mu_B}{d n_B} \qquad \text{with} \quad \mu_B = \frac{d \varepsilon}{d n_B},
\end{equation} 
for our preferred scenario. 
There we see, as in Ref. \cite{Fujimoto:2023mzy}, a singular behavior at the density, $n_{onset}$,  where quark-distribution at zero momentum becomes unity, $f_Q(q=0)=1$. We consider this singularity as an artifact of the specific choice of implementing the quark distribution inside a nucleon, given by Eq.~\eqref{eq:quark_dist}. 
Interestingly, the onset density, $n_{\rm onset} \simeq 0.53 n_0$, is very similar (but not identical) in symmetric and neutron matter.
For the same scenario, in Fig. \ref{fig:dist} we show the momentum distribution of the baryons (nucleon) at ground state nuclear matter density, $n_0$. The shell structure in the momentum distribution is clearly visible. For momenta $k < k_{\rm bu} \simeq 180 \,\rm MeV$ the momentum distribution is suppressed,  $f_B(k) = 1/N_c^3 \simeq 0.037 \ll 1$. As a consequence, the maximum (or Fermi) momentum of the nucleons is pushed to $k_{F}=288 \,\rm MeV$, considerably larger than that of an ordinary Fermi gas for the same density, $k_{F,\rm regular}\simeq 260 \, \rm MeV$.  

As we will later see, the formation of a hole at low momentum in the Fermi momentum distribution should be seen in quasi-elastic electron scattering data from nuclei.  We will later see that there is evidence for such a hole, but at a lower value of momentum $k_{\rm hole}  \sim 100$~MeV. Reconciling this difference might be perhaps due to modifying the probability distribution $\phi$ if the IdylliQ model, perhaps properly including the effects of a density dependent nucleon mass, or perhaps some residual effect due to vector meson exchange.

\subsection{The Effect of a Density-Dependent Nucleon Mass}

With the effect of the density dependence of the nucleon  mass $m_N^*$ included, 
 the energy density reads
 \begin{equation}
 \label{eq:edens:meff}
 \varepsilon(n_B) = \varepsilon_K^* + \frac{g_s^2}{2m_{\sigma}^2} (n_{s}^*)^2 + \varepsilon_{\rm exch}^{\sigma *} + \varepsilon_{\rm exch}^{\pi *},
\end{equation}
where terms $\varepsilon_K^*$, $n_s^*$, $\varepsilon_{\rm exch}^{\sigma *}$, $\varepsilon_{\rm exch}^{\pi *}$ are given by Eqs.~\eqref{eq:eK}, \eqref{eq:rhosig}, \eqref{eq:exsig}, \eqref{eq:expi}, respectively, where we substitute vacuum nucleon mass $m_N$ by the effective mass $m_N^*$. The effective mass $m_N^*$ is determined on a mean-field level from the equation
\begin{equation}
m_N^* = m_N - \frac{g_s^2}{m_\sigma^2} n_s^*.
\end{equation}
which minimizes the mean-field level energy density $\varepsilon(\rho_B) = \varepsilon_K^* + \frac{m_\sigma^2}{2 g_s^2} (m_N - m_N^*)^2$  with respect to $m_N^*$.
\footnote{ Alternatively, we have also considered including $\varepsilon_{\rm exch}^{\sigma*}$ and $\varepsilon_{\rm exch}^{\pi*}$ terms when determining the effective mass through energy minimization, generally obtaining a stiffer EoS but with same qualitative features.}

\begin{figure}[t]
    \includegraphics[width=.32\textwidth]{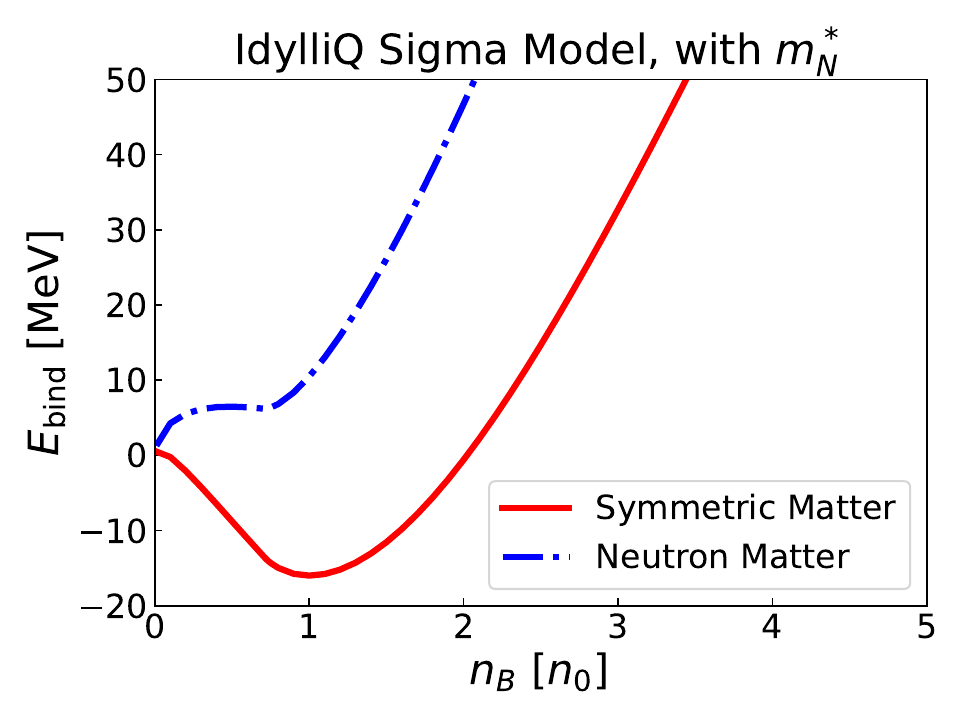}
    \includegraphics[width=.32\textwidth]{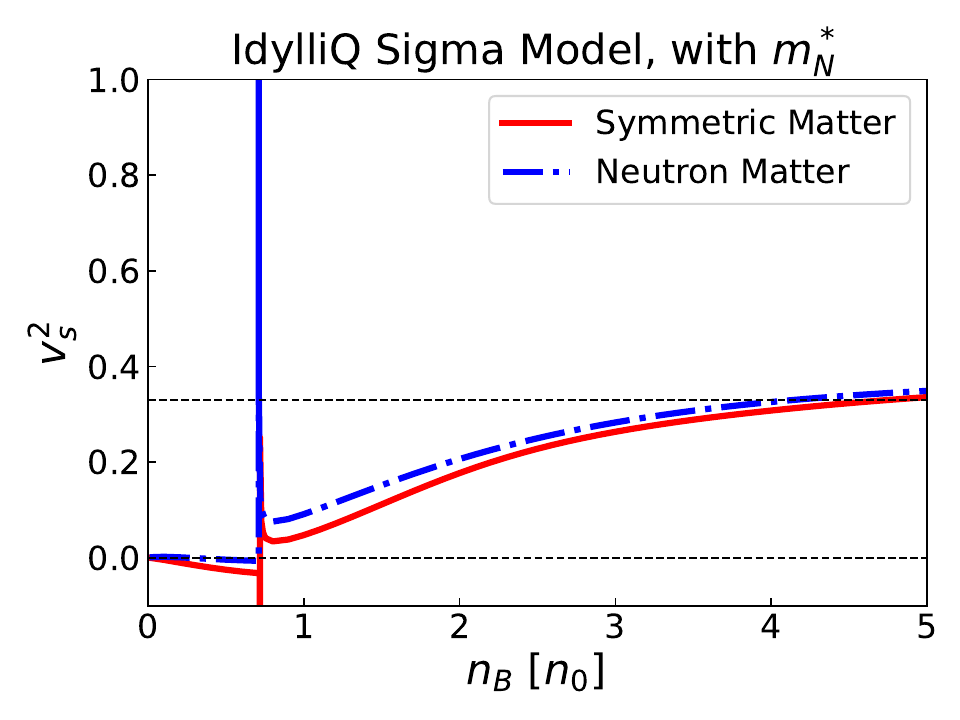}
       \includegraphics[width=.32\textwidth]{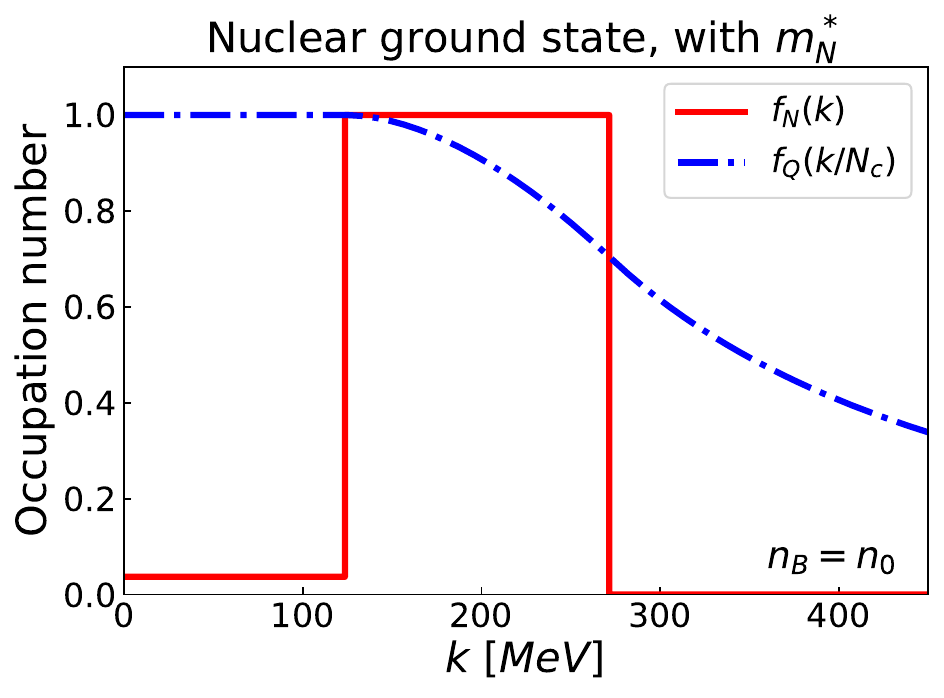}
    \caption{
    Binding energy~(left panel), the squared sound velocity~(middle panel) of symmetric nuclear~(solid red) and pure neutron~(dash-dotted blue) matter, and the occupation numbers~(right panel) of nucleons~(solid red) and quarks~(dash-dotted blue) as a function of the momentum $k$ in the nuclear ground state. 
    The figure depicts calculations within the IdylliQ Sigma Model with a density-dependent mass $m_N^*$ and $m_\sigma = 0.9$~GeV, scalar and pion exchange energy terms included.
    }
    \label{fig:EoSmeff}
\end{figure}

The results with a density-dependent mass are shown in Fig. \ref{fig:EoSmeff} for the binding energy, sound velocity, and the occupation numbers of quarks and nucleons at $n_B = n_0$.
The calculation was performed for heavy scalar meson mass $m_\sigma = 900$~MeV/$c^2$, and includes both the scalar and pion exchange energy terms, i.e. it corresponds to our preferred scenario from the previous subsection but with the additional effect of density-dependent nucleon mass included.
The results are qualitatively similar to the fixed-mass calculation but with notable quantitative differences.
The Quarkyonic onset density is $n_{\rm onset} \simeq 0.71 n_0$, for both symmetric and pure neutron matter, which is considerably closer to $n_0$ than $n_{\rm onset} \simeq 0.53n_0$ for fixed-mass calculation.
Consequently, the stiffening of the equation of state near $n_B \sim n_0$ is more rapid, leading to a large value of the compressibility $K_0 \simeq 395$~MeV. 
The lower momentum of the shell at $n_B = n_0$ is $k_{\rm bu} \simeq 124$~MeV, considerably lower than $k_{\rm bu} \simeq 180$~MeV in the fixed mass calculation.
Also, the sound velocity is observed to exceed 1/3 at the highest densities considered~($n_B \sim 5 n_0$) for neutron matter. 
We have performed density-dependent mass calculation also for other values of $m_\sigma$ and for different scenarios regarding the inclusion of exchange energy terms.
Our description of nuclear matter can certainly be improved. For example the large value of the compressibility, which is typical for the Walecka model, can be be reduced by introducing non-linear terms in for the $\sigma$-field \cite{Waldhauser:1987ed}, which is beyond the scope of the present work. However, qualitatively we observe an anti-correlation between the values of the lower momentum of the shell $k_{\rm bu}$ and compressibility $K_0$.

To conclude this section, we have shown that the IdylliQ model can reproduce the main feature of nuclear matter. If correct, this would imply that it should be realized in ordinary nuclei. In particular, the IdylliQ model predicts a hole in the nucleon momentum distribution, which should be accessible to electron scattering experiments on nuclei. To which extent this is the case we will discuss next.

\section{Electron Scattering}
\label{sec:elscat}
 
The previous sections present a startling new view of the nucleon momentum distribution of nuclear matter, indicating a strong depletion of the Fermi sea of nucleons at low momenta, $k \lesssim 120-180$~MeV. 
Clearly, it is necessary to check if such a distribution is consistent with known experimental data.
 
Quasi-elastic $(e,e')$ electron scattering from nuclei has provided some of the best information about nucleon momentum distributions~\cite{Sick:1970ma,Day:1987az,Day:1989nw,Moniz:1969sr,Moniz:1971mt,Benhar:1991af,Benhar:2006wy}.  
We can understand this qualitatively. 
In the one-photon-exchange approximation, an electron emits a photon of four-momentum, $q=(\omega,{\bf q})$ that hits a nucleon and knocks it out of the nucleus. 
If the nucleon is at rest and on its mass-shell, it will emerge with an energy $-q^2/(2m_N) = Q^2/(2m_N)$. 
So at fixed spatial momentum transfer, there will be a sharp peak at a fixed energy corresponding to this process.  
Taking the initial non-zero momentum of the nucleon into account, this peak is spread out by the Fermi motion and also shifted by an average nucleon interaction energy~\cite{Moniz:1969sr}.
On the other hand, if the nucleon momentum distribution is as given by Quarkyonic matter, then there will be a depletion of nucleons at low momentum relative to that of the elastic peak.  
This will affect both the shape and the height of the momentum distribution at small momenta in the kinematic region of the quasi-elastic peak of electron-nucleus scattering.
 
We compare the effects of using the standard Fermi distribution with that of the Quarkyonic distribution using the following formalism, consistent with the notation of Ref.~\cite{Benhar:2006wy}. 
The $(e,e') $ cross section is given by
\bea {d\s\over d\Omega dE_e}= {d\s_M\over d\Omega}[W_2 +2W_1 \tan^2(\th/2)],\label{1}
\eea
where the Mott cross section, ${d\s_M\over d\Omega}$ is given by~\cite{Thomas:2001kw}
\bea {d\s_M\over d\Omega} ={\alpha^2\cos^2({\th/2})\over 4E_e^2\sin^4(\th/2)\, (1+2E_e/m_N \sin^2\th/2)}
\eea
The cross-section is proportional to $L_{\m\n}W^{\m\n}$ with $L_{\m\n}$ the lepton tensor and $W^{\m\n} $ the target tensor:
\bea
W^{\m\n}=W_1\left(g^{\m\n}-{q^\m q^\n \over q^2}\right)+
{W_2\over M_A^2}
\left(P^\m - {q\cdot P\over q^2} q^\m\right)
\left(P^\n-{q\cdot P\over q^2}q^\n \right),
\label{3}
\eea
with the lab-frame  nuclear four-vector, $P^\m=(M_A,{\bf0})$.
In the Fermi gas model, the nucleons are on their mass shell in both the initial and final state. 
Then $W^{\m\n}$ from a nuclear target is given by
\bea 
W_1\left(g^{\m\n}-{q^\m q^\n\over q^2}\right)  +{W_2\over M_A^2}\left(P^\m-{q\cdot P\over q^2}q^\m\right)\left(P^\n-{q\cdot P\over q^2}q^\n\right) 
=
\int_k \rho(k) w_N^{\m\n}
\label{4}
\eea
where we introduced a short-hand
\begin{equation}
\int_k\equiv2\sum_{N=n,p}\int {\frac{d^3k}{(2\pi)^3} \d\left(\w + E_k - \sqrt{k^2 + 2 \bfk\cdot\bfq+\bfq^2 + m_N^2} \right)}.
\end{equation}

The nucleon tensor, $w_N^{\m\n}$, is given by
\bea w_N^{\m\n}(k,q) & = &w_1^N(g^{\m\n}-{q^\m q^\n\over q^2})
+{w_2^N\over m_N^2}(k^\m-{q\cdot k\over q^2}q^\m)(k^\n-{q\cdot k\over q^2}q^\n)\label{5}\\
w_1 & = &{Q^2\over 4m_N^2}G_M^2(Q^2),\\
w_2 & = & {1\over 1+{Q^2\over 4m_N^2}}[G_E^2(Q^2) +{Q^2\over 4m_N^2}G_M^2(Q^2)]
,\eea
with 
$G_{E,M}(Q^2)$ being  the elastic electric and magnetic Sachs form factors and   $k=(E_k,\bfk),\, E_k=\sqrt{k^2+m_N^2}$ and $q\cdot k=-q^2/2=Q^2/2$.

The normalization of $\rho(k) $ is
\bea A=2\int {d^3k\over (2\pi)^3}\rho(k)=2 V \int {d^3k\over (2\pi)^3} [f_p(k) + f_n(k)]\label{Adef}\eea
with $\rho(k)=V [f_p(k) + f_n(k)] $ being proportional to the nucleon momentum space density\footnote{Note that the volume $V$ drops out in the final, normalized, expression}.
For Quarkyonic matter the occupation numbers $f_{p,n}(k)$ are given by Eq.~\eqref{eq:piecewise} while for regular nuclear matter they are given by the usual Fermi-distributions.

We will compare our calculations with the experimental data on nuclear matter cross sections obtained by Day et al~\cite{Day:1989nw}, $\Sigma\equiv {1\over A}{d\s\over d\Omega dE_e}$, in the limit $A\to\infty$.

Next we need to identify $W_{1,2}$ in terms of $w_{1,2}$. This is done by multiplying \eq{4}  by $g_{\m\n}$ and then multiplying \eq{4} by $P_\m P_\n$ to get two equations and two unknowns.  
\bea 
W_1=\int_k \left[w_1^N+{w_2^N\over 2}\left(1+\frac{Q^2}{4m_N^2}-f (\w,k) \frac{Q^2}{\bfq^2}\right)\right]. \eea

and
\bea W_2=\int_k \left\{ w_2^N \left[{3\over2}f (\w,k){Q^4\over \bfq^4}-{1\over2}{Q^2\over \bfq^2}\left(1+\frac{Q^2}{4m_N^2} \right) \right] \right\}
,
\eea
where $f(\w,k)=(E_k+ \w/2)^2/m_N^2$.

In the following, we compare with the 500 MeV electron scattering data of \cite{Day:1989nw}.
The form factors are well described by the simple dipole forms used in ~\cite{Moniz:1969sr}.
Our calculations ignore the effects of final state interactions. These are about 5\% for the data under discussion~\cite{Horikawa:1980cg}, which is about the size of the normalization uncertainty, but are larger at higher energies~\cite{Benhar:1991af}. Therefore we focus our discussion on the 500 MeV data.

\begin{figure}[t]
    \includegraphics[width=.65\textwidth]{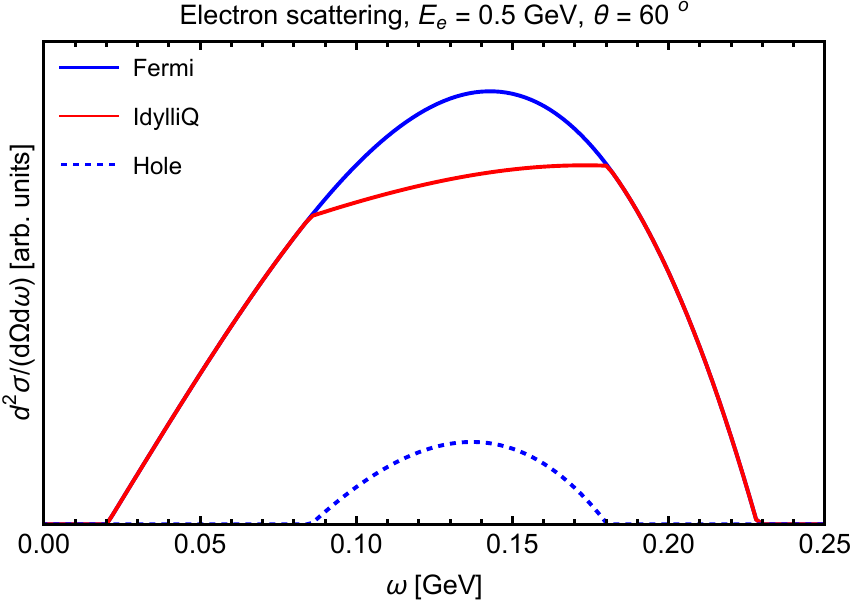}
    \caption{
    Quasi-elastic electron scattering cross section
    in nuclear matter calculated using the standard Fermi distribution with $k_F = 271.5$~MeV~(solid blue line), and the IdylliQ distribution where the nucleon states at momenta at $k < k_{\rm sh} = 123.6$~MeV are suppressed by factor $N_c^{-3}$~(solid red line).
    The dashed blue line shows the difference between the two, which corresponds to the contribution of the hole in the Fermi sea.
    }
    \label{fig:elscat:effect}
\end{figure}

First, we discuss how the presence of the hole in the Fermi sea affects electron scattering cross section qualitatively.
This effect is shown in Fig.~\ref{fig:elscat:effect}.
The solid blue line depicts the cross section, here in arbitrary units, for ordinary Fermi distribution with $k_F = 271.5$~MeV.
The solid red line shows the results for IdylliQ type distribution, there all momentum states at $k < k_{\rm sh} = 123.6$~MeV are suppressed by a factor $N_c^{-3}$ as per Eq.~\eqref{eq:piecewise}.
The dashed blue line shows the difference between the two scenarios, i.e., it is the contribution of the hole.
The hole modifies the shape of the cross section around the peak, by generally making it flatter and visibly asymmetric.
The larger the hole is, the stronger the effect.

\begin{figure}[t]
    \includegraphics[width=.90\textwidth]{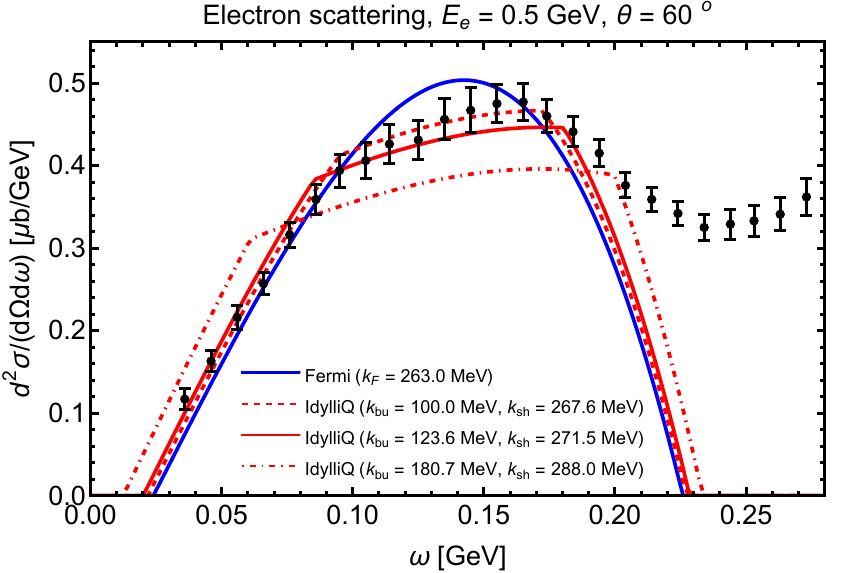}
   
    \caption{
    Quasi-elastic electron scattering cross section in nuclear matter calculated using the standard Fermi distribution~(blue line), and for three cases of the IdylliQ distribution with the varying size of the quark Fermi sea: $k_{\rm bu} = 100.0$~MeV~(dashed red), $k_{\rm bu} = 123.6$~MeV~(solid red), and $k_{\rm bu} = 180.7$~MeV~(dash-dotted red). 
    The values of upper Fermi momentum $k_F/k_{\rm sh}$ are fixed in each case to reproduce normal nuclear density.
    The black points depict the experimental data from Day et al.~\cite{Day:1989nw}.
    }
    \label{fig:elscat:withdata}
\end{figure}

In Figure~\ref{fig:elscat:withdata} we compare our calculations to the experimental data~\cite{Day:1989nw}.
Calculations are done for the standard Fermi distribution~(blue line), and for three cases of the IdylliQ distribution where we vary the size of the quark Fermi sea: $k_{\rm bu} = 180.7$~MeV~(preferred scenario without density-dependent mass), $k_{\rm bu} = 123.6$~MeV~(preferred scenario with density-dependent mass), and $k_{\rm bu} = 100.0$~MeV.
In all cases, the value of $k_{\rm sh}$ is chosen so that one reproduces the normal nuclear density of $n_0 = 0.16$~fm$^{-3}$. 
For comparisons with the experiment, we used an average nucleon interaction energy of 30~MeV. 

Calculations describe the experimental data overall well for energies $\omega \lesssim 0.19$~GeV that are appropriate for quasi-elastic scattering.
Looking more into quantitative details, one can see indications in the experimental data for an asymmetric shape of the peak, with maximum corresponding to $\omega \sim 160-170$~MeV, while the standard Fermi distribution predicts a largely symmetric peak centered around $\omega = 140$~MeV.
The IdylliQ calculations with $k_{\rm sh} = 120$~MeV and, more so, $k_{\rm sh} = 100$~MeV, appear to give the best description of the features seen in experimental data.
On the other hand, for $k_{\rm bu} = 180.7$~MeV, the peak becomes too broad and is not supported by the data.

Overall, we can conclude that experimental data do not rule out, and to some extent even indicate, the presence of a strong depletion of nucleons in nuclear matter at momenta $k \lesssim 120$~MeV.

\section{Summary}

\begin{itemize}
    \item Using reasonable parameters for the sigma and pion interactions, and a density-independent nucleon mass, we find that the properties of ground-state nuclear matter can reasonably be described. 
    The compressibility of nuclear matter ranges from $180 \rm \, MeV$ to $250 \rm \, MeV$, which is comparable with values extracted from experiments \cite{Danielewicz:2002pu}. 
    Also, the predicted symmetry energy at $n_0$ is in an acceptable range with $29 \, \rm MeV \lesssim S_0 \lesssim 32 \, \rm MeV$ \cite{Baldo:2016jhp}. 
    This is achieved {\it without} the need for any repulsive vector interaction. 
    Instead, the needed repulsion originates from the Pauli exclusion in the quark sector. 
   
    \item In order to achieve saturation of nuclear matter at the correct ground state density, $n_B=n_0$, the density where the shell structure in the baryon momentum distribution emerges needs to be below saturation density, $n_{\rm onset}<n_0$. 
    Specifically, we find 
    $n_{\rm onset} \simeq 0.53 n_0$ for the scenario considered here. 
    Thus, our model/theory predicts a shell structure in the momentum distribution or equivalently saturated quark distribution at low momenta which should be present in ordinary nuclei. As shown,  this prediction is not inconsistent with data from quasi-elastic electron scattering on nuclei.
    
    \item We find that the speed of sound exhibits non-analytic behavior at the onset density, $n_{\rm onset}$. We consider this an artifact of the present (analytic) implementation of the IdylliQ model. This singular behavior occurs in a very narrow range of density,
    which shrinks to zero as $N_c \to \infty$.  An explicit solution for this problem has not yet been presented.
    
    \item We find that neutron matter is unbound, at least if sigma and/or pion exchange interactions are taken into account. In the range of interest, we find that the speed of sound is below $1/3$  for nuclear matter, which is a bit too soft for describing neutron stars. 
    
    \item  If we allow the nucleon mass to be density-dependent, the equation of state becomes stiffer, and the lower momentum for the nucleonic shell decreases.  However, the compressibility becomes larger than the measured values.

    \item The IdylliQ sigma model description of nuclear matter presented here substantially differs from most other frameworks used to describe nuclear matter at normal densities. There, the relation between the Fermi momentum and number density is the same, or very close to that of the free gas, 
    whereas the momentum distribution is a theta function, $f_B(k) = \Theta(k_F-k)$.
    In contrast, in the IdylliQ sigma model, the baryon occupation number is strongly suppressed at low momenta already at normal nuclear densities, $n_B \simeq n_0$~(Fig.~\ref{fig:dist}).
    Our framework does show some similarity to the quantum van der Waals~(QvdW) theory of nuclear matter~\cite{Vovchenko:2015vxa}. In the QvdW model, the attractive interactions are described by a mean field, which is equivalent to the scalar $\sigma$ field in the non-relativistic limit. On the other hand, the repulsion is modeled by the excluded volume correction, which modifies the relation between the nucleon density and the Fermi momentum and thus schematically models the Pauli exclusion principle among constituent quarks~\cite{Poberezhnyuk:2023rct}.
    
    \item  Of course, this very simple computation leaves out much physics associated with interactions.  Nevertheless, it has the correct qualitative and semi-quantitative features to describe nuclear matter
    \item  The most direct test of the idea that nuclear matter is Quarkyonic occurs via quasi-elastic electron $(e,e')$ scattering. Our calculations for scattering at 500 MeV are, somewhat surprisingly, in reasonable agreement with the data. This energy is chosen because the effects of final state interactions are small. Copious data exists at other energies,  and could be addressed when such interactions are included.
      
\end{itemize}
 
\section{Conclusion}
\label{sec:summary} 

The current state of the model is primitive. The model may be refined in several ways ranging from different choices of the quark distribution in the nucleon to possible higher order interaction effects. Also, of interest may be the properties of this model at finite temperature and for systems with strange degrees of freedom. Computations of $(e,e')$ cross sections could be made at other kinematics.

Despite these caveats, it is fair to say that we have demonstrated that the IdylliQ model for Quarkyonic matter is able to describe ground state nuclear matter as well as low density neutron matter. The essential new element is that in this approach the necessary repulsion is provided by the Pauli exclusion principle in the quark sector. The present basic implementation of the IdylliQ model captures what we believe the essential feature of Quarkyonic matter, namely that the low momentum nucleon states are depleted. Or in other words, the nucleon Fermi distribution has a "hole" for momenta $k \lesssim 120 \,\rm MeV$. Remarkably, such a hole in the Fermi distribution is not only consistent with quasi-elastic scattering data but there may actually be even an indication for it in the shape of the data.

\section*{Acknowledgements}
L.M. acknowledges the support by the US Department of Energy under contract number DE-FG02-00ER41132.
V.K.  would like to thank the Institute for Nuclear Theory at the University of Washington for the kind hospitality and stimulating research environment.   
V.V. thanks the Nuclear Theory Group at Lawrence Berkeley National Laboratory, where he has done part of this work, for the kind hospitality.
This work has been supported by the U.S. Department of Energy, Office of Science, Office of Nuclear Physics, under contract number DE-AC02-05CH11231, by the INT's U.S. Department of Energy grant No. DE-FG02-00ER41132 and the DOE grant No. DE-FG02-97ER41014. We thank Douglas Higinbotham for useful discussions.

\bibliography{paper}

\end{document}